\documentclass{article}
\usepackage[utf8]{inputenc}
\usepackage{float}
\usepackage{amsfonts}
\usepackage{graphicx}
\usepackage{color}
\usepackage{subcaption}
\usepackage{xspace}
\usepackage{xifthen}
\usepackage{multicol}
\usepackage{mathtools}
\usepackage{algorithm,algorithmicx}
\usepackage{bbm}
\usepackage{wasysym}
\usepackage[makeroom]{cancel}
\usepackage{amsmath}
\usepackage[english]{babel}
\usepackage{graphicx}
\usepackage{fontenc}

\usepackage{amssymb}
\usepackage{dsfont}
\usepackage{amsthm}
\usepackage{amssymb}
\usepackage{placeins}
\usepackage[toc,page]{appendix}
\usepackage{xcolor}
\usepackage{enumerate}
\usepackage{tabularx}
\usepackage{algcompatible}
\usepackage{hyperref}
\usepackage[font=small,labelfont=bf]{caption}
\usepackage{csquotes}
\usepackage{ mathrsfs }

\newcommand{\blind}{0}
\usepackage{listings}

\usepackage{xcolor}

\definecolor{codegreen}{rgb}{0,0.6,0}
\definecolor{codegray}{rgb}{0.5,0.5,0.5}
\definecolor{codepurple}{rgb}{0.58,0,0.82}
\definecolor{backcolour}{rgb}{0.95,0.95,0.92}

\lstdefinestyle{mystyle}{
    backgroundcolor=\color{backcolour},   
    commentstyle=\color{codegreen},
    keywordstyle=\color{magenta},
    numberstyle=\tiny\color{codegray},
    stringstyle=\color{codepurple},
    basicstyle=\ttfamily\footnotesize,
    breakatwhitespace=false,         
    breaklines=true,                 
    captionpos=b,                    
    keepspaces=true,                 
    numbers=left,                    
    numbersep=5pt,                  
    showspaces=false,                
    showstringspaces=false,
    showtabs=false,                  
    tabsize=2
}

\begin{document}

\def\spacingset#1{\renewcommand{\baselinestretch}%
{#1}\small\normalsize} \spacingset{1}

\if0\blind
{
  \title{\bf 
  FunQuant: A R package to perform quantization in the context of rare events and time-consuming simulations}
  \author{Charlie Sire \\
    \small IRSN, BRGM, Mines Saint-Etienne, Univ. Clermont Auvergne, CNRS, UMR 6158 LIMOS\\
    and \\
    Yann Richet \\
    \small IRSN \\
        and\\
    Rodolphe Le Riche \\
    \small CNRS, Mines Saint-Etienne, Univ. Clermont Auvergne, \small UMR 6158 LIMOS\\
    and\\
     Didier Rullière \\ 
    \small Mines Saint-Etienne, Univ. Clermont Auvergne, CNRS, UMR 6158 LIMOS\\
    and \\
    Jérémy Rohmer \\
    \small BRGM\\
    and \\
     Lucie Pheulpin \\
    \small IRSN 
    }
    \date{}
  \maketitle
} \fi
\if1\blind
{
  \bigskip
  \bigskip
  \bigskip
  \begin{center}
    {\LARGE\bf Title}
\end{center}
  \medskip
} \fi

\bigskip

\section{Summary} 

Quantization summarizes continuous distributions by calculating a discrete approximation~\cite{Pages}. Among the widely adopted methods for data quantization is Lloyd's algorithm, which partitions the space into Voronoï cells, that can be seen as clusters, and constructs a discrete distribution based on their centroids and probabilistic masses. Lloyd's algorithm estimates the optimal centroids in a minimal expected distance sense~\cite{Bock}, but this approach poses significant challenges in scenarios where data evaluation is costly, and relates to rare events. Then, the single cluster associated to no event takes the majority of the probability mass. In this context, a metamodel is required~\cite{Friedman} and adapted sampling methods are necessary to increase the precision of the computations on the rare clusters.

\section{Statement of need}

{\ttfamily FunQuant} is a R package that has been specifically developed for carrying out quantization in the context of rare events. While several packages facilitate straightforward implementations of the Lloyd's algorithm, they lack the specific specification of any probabilistic factors, treating all data points equally in terms of weighting. Conversely,  {\ttfamily FunQuant} considers probabilistic weights based on the Importance Sampling formulation~\cite{Paananen} to handle the problem of rare event. To be more precise, when $X$ and $Y$ are the random vectors of inputs and outputs of a computer code,  the quantization of $Y(X)$ is performed by estimating the centroid of a given cluster $C$ with the following formula,

$$\frac{\frac{1}{n} \sum^{n}_{k=1} Y(\tilde{X}_{k})\mathds{1}_{Y(\tilde{X}_{k})\in C}\frac{f_{X}(\tilde{X}_k)}{g(\tilde{X}_{k})}}{\frac{1}{n} \sum^{n}_{k=1} \mathds{1}_{Y(\tilde{X}_k)\in C} \frac{f_{X}(\tilde{X}_k)}{g(\tilde{X}_{k})}},$$
where $f_{X}$ is the known density function of the inputs $X$, and $(\tilde{X}_k)^{n}_{k=1}$ i.i.d. random variables of density function $g$.
Importance Sampling is employed with the aim of reducing the variance of the estimators of the centroids when compared to classical Monte Carlo methods. {\ttfamily FunQuant} provides various approaches for implementing these estimators, depending on the sampling density $g$. The simplest method involves using the same function $g$ for each iteration and every cluster, which is straightforward to work with and still yields significant variance reductions. More advanced implementations enable the adaptation of the sampling density for each cluster at every iteration.

In addition, {\ttfamily FunQuant} is designed to mitigate the computational burden associated with the evaluation of costly data. While users have the flexibility to use their own metamodels to generate additional data, {\ttfamily FunQuant} offers several functions tailored specifically for spatial outputs such as maps. This metamodel relies on Functional Principal Component Analysis and Gaussian Processes, based on the work of \cite{Perrin}, adapted with the {\ttfamily rlibkriging} R package~\cite{rlib}. {\ttfamily FunQuant} assists users in the fine-tuning of its hyperparameters for a quantization task, by providing a set of relevant performance metrics.

Additional theoretical information can be found in \cite{sire}. The paper provides a comprehensive exploration of the application of {\ttfamily FunQuant} to the quantization of flooding maps.

\section{Illustrative example}

\begin{figure}[H]
    \centering
    \includegraphics[width=0.7\textwidth]{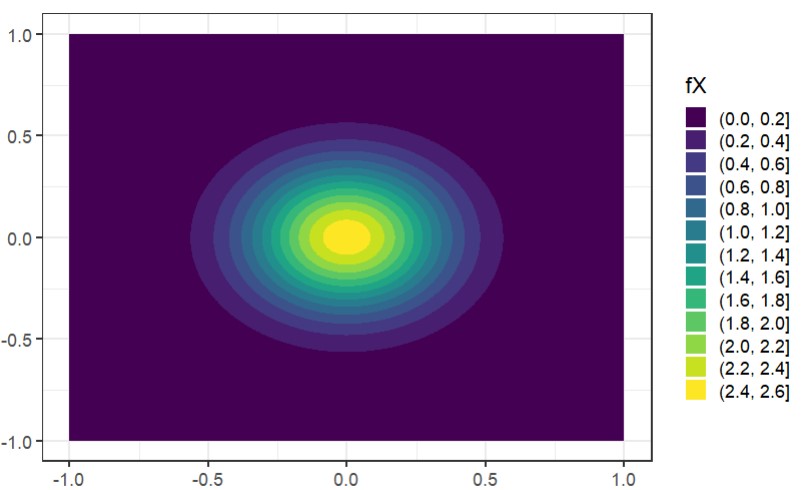}
    \caption{Density function $f_{X}$.}
    \label{fx}
\end{figure}

We consider $X = (X_{1},X_{2}) \in \mathbb{R}^2$ a random input of a computer code $H$, with
$$\left\{
    \begin{array}{ll}
        X_{i} \sim \mathcal{N}_{t}(0,0.25^2, -1, 1), i=1,2 \\
        X_{1} \text{ and }X_{2}\text{ independent}
    \end{array}
\right.$$

where $\mathcal{N}_{t}(\mu,\sigma^2, a, b)$ is the Gaussian distribution of mean $\mu$, variance $\sigma^2$, truncated between $a$ and $b$. The density function of $X$, denoted $f_{X}$, is represented in Figure~\ref{fx}.

The computer code $H$ is defined with 
$$H(x) = \left\{
    \begin{array}{ll}
        (0,0) \text{ if } \lvert x_{1}\rvert \leq \alpha \\
        (\lvert x_{1} \rvert - \alpha, \lvert x_{2} \rvert) \text{ otherwise.}
    \end{array}
\right.$$

with $\alpha$ such that $P(H(X) = (0,0)) = 0.99.$

\begin{figure}[H]
    \centering
    \includegraphics[width=0.7\textwidth]{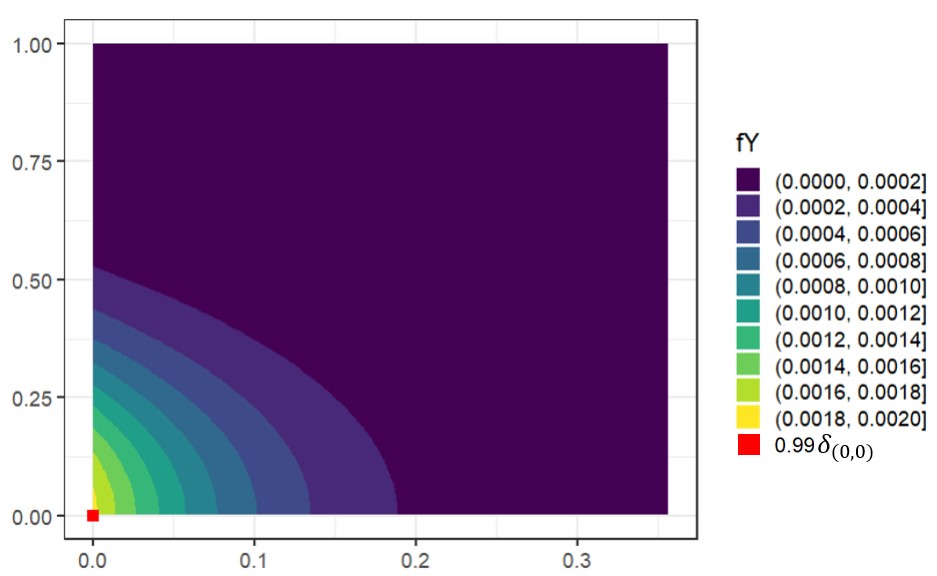}
    \caption{Density function $f_{Y}$.}
    \label{fy}
\end{figure}

The density $f_{Y}$ of the output $Y = H(X)$ is represented in Figure~\ref{fy}.

$99\%$ of the probability mass is concentrated at $(0,0)$. We want to quantize $Y(X)$.

\begin{figure}[H]
    \centering
    \includegraphics[width=0.8\textwidth]{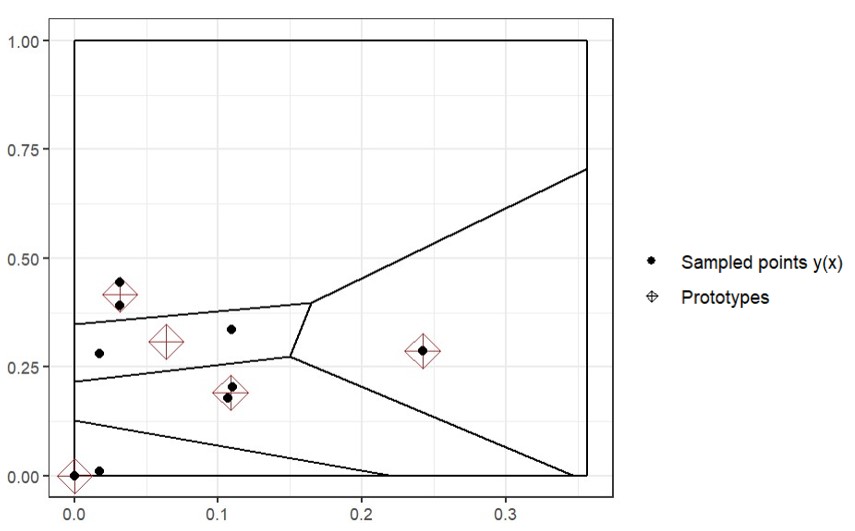}
    \caption{Sampling and quantization with classical Lloyd.}
    \label{kmeans_quanti}
\end{figure}

 If the classical Lloyd's algorithm is run with a budget of $1000$ points, it leads to the outcome illustrated in Figure~\ref{kmeans_quanti}, with only a few sampled points not equal to $(0,0)$. Then, the centroids of the Voronoi cells that do not contain $(0,0)$ are computed with a very small number of points, leading to a very high variance.

The {\ttfamily FunQuant} package allows to adapt the sampling by introducing a random variable $\tilde{X}$ of density $g$, and considering the probabilistic weights of each sample, with are the ratio $\frac{f_{X}}{g}$.

A possible function $g$ is $g(x) = \frac{1}{4}\mathds{1}_{[-1,1]^2}(x)$, corresponding to a uniform distribution in $[-1,1]^2$.
\lstset{style=mystyle}

\begin{lstlisting}[language = R]
fX = function(x){
  return(
    dtruncnorm(x = x[1],mean = 0,sd = sd1,a=-1, b=1)*dtruncnorm(x = x[2],mean = 0,sd = sd2,a=-1, b=1))
}

g = function(x){
  if(sum((x>-1)*(x<1))==2){return(1/4)}
  else{return(0)}
}

sample_g = function(n){cbind(runif(n,-1,1), runif(n,-1,1))
}

inputs = sample_g(1000)
outputs = t(apply(inputs,1,Y))
density_ratio = compute_density_ratio(f = fX, 
                                      g = g, 
                                      inputs = inputs)
                                    
res_proto = find_prototypes(data = t(outputs),
                            nb_cells = 5,
                            multistart = 3,
                            density_ratio = density_ratio)
\end{lstlisting}

\begin{figure}[H]
    \centering
    \includegraphics[width=0.7\textwidth]{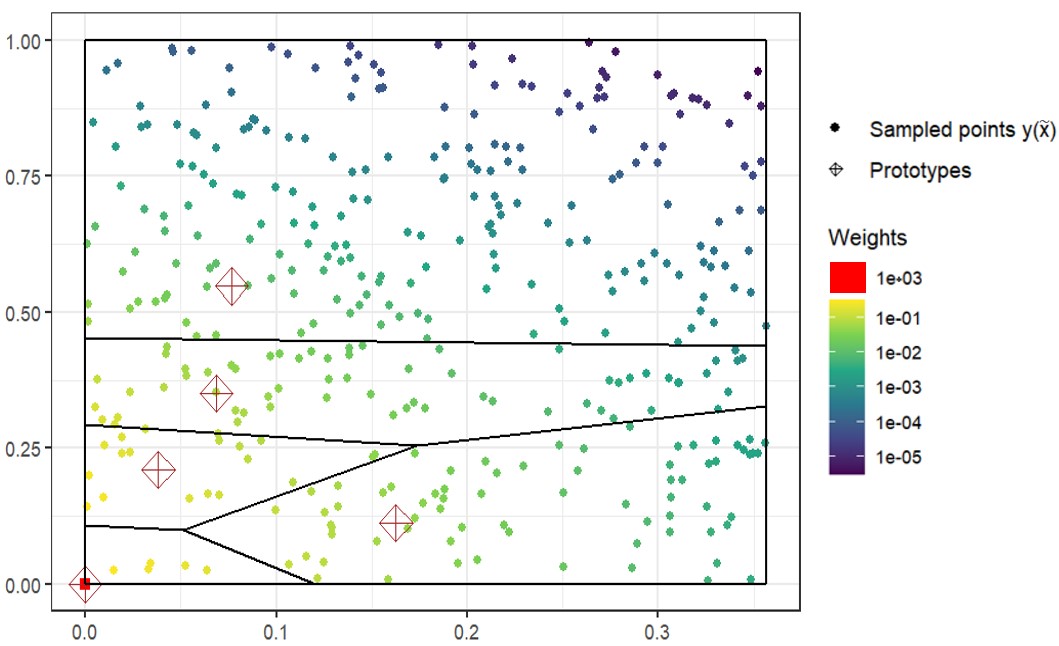}
    \caption{Sampling and quantization with importance sampling weights.}
    \label{is_quanti}
\end{figure}

Figure~\ref{is_quanti} shows the sampled points $Y(\tilde{X}_{k})$, their associated probabilistic weights, and the obtained prototypes. It clearly appears that this sampling brings more information about each Voronoi cells. 

{\ttfamily FunQuant} allows to estimate the standard deviations of the two coordinates of the estimators of the centroids for each Voronoi cell, highlighting the variance reduction obtained with the adapted sampling for the cells that do not contain $(0,0)$.

\begin{lstlisting}[language = R]
large_inputs = sample_fX(10^5)
large_outputs = apply(large_inputs,1, Y)
std_centroid_kmeans = std_centroid(
              data = large_outputs, 
              prototypes_list = list(protos_kmeans),
              cells = 1:5, 
              nv = 1000)

std_centroid_kmeans #the cells are ordered by the increasing coordinate x of their centroid

# std centroid returns a list of lists: for each tested set of prototypes (here only one set is tested), a list of the estimated standard deviations is provided, each element of this list is associated to a cell

\end{lstlisting}

\begin{lstlisting}[language = R]
[[1]]
[[1]][[1]]
[1] 0.0001193543 0.0001012730
  
[[1]][[2]]
[1] 0.04884616 0.07905258
[[1]][[3]]
[1] 0.03006552 0.02934998

[[1]][[4]]
[1] 0.03214239 0.02801202
 
[[1]][[5]]
[1] 0.06158175 0.12912278
\end{lstlisting}

\begin{lstlisting}[language = R]
large_inputs_is = sample_g(10^5)
large_outputs_is = apply(large_inputs_is,1, Y)
std_centroid_FunQuant = std_centroid(
              data = large_outputs_is, 
              prototypes_list = list(protos_FunQuant),
              cells = 1:5, 
              nv = 1000)

std_centroid_FunQuant #the cells are ordered by the increasing coordinate x of their centroid
\end{lstlisting}

\begin{lstlisting}[language = R]
[[1]]
[[1]][[1]]
[1] 0.0002358303 0.0002390596

[[1]][[2]]
[1] 0.00901367 0.01033904

[[1]][[3]]
[1] 0.012857642 0.006439004

[[1]][[4]]
[1] 0.00726317 0.01139948

[[1]][[5]]
[1] 0.009168924 0.009620646
\end{lstlisting}


This example remains basic. Advanced computations of the centroids with tailored density functions $g$ can be performed. {\ttfamily FunQuant} was built to tackle industrial problems with large amounts of data, and comes with additional features such as the possibility to split the computations into different batches. 

\section{Acknowledgments}

This research was conducted with the support of the consortium in
Applied Mathematics CIROQUO, gathering partners in technological and
academia towards the development of advanced methods for Computer
Experiments.

\bibliographystyle{plain}
\bibliography{biblio_manuscrit.bib}

\begin{thebibliography}{1}

\bibitem{Bock}
Hans-Hermann Bock.
\newblock Origins and extensions of the k-means algorithm in cluster analysis.
\newblock {\em Journal Électronique d’Histoire des Probabilités et de la
  Statistique [electronic only]}, 4:Article 14, 01 2008.

\bibitem{Friedman}
Linda~Weiser Friedman.
\newblock {\em The Simulation Model and Metamodel}, pages 13--31.
\newblock Springer US, Boston, MA, 1996.

\bibitem{rlib}
Pascal Havé, Yann Richet, Yves Deville, Conrad Sanderson, Colin Fang, Ciyou
  Zhu, Richard Byrd, Jorge Nocedal, and Jose~Luis Morales.
\newblock rlibkriging: Kriging models using the 'libkriging' library, 2022.

\bibitem{Paananen}
Topi Paananen, Juho Piironen, Paul-Christian Bürkner, and Aki Vehtari.
\newblock Implicitly adaptive importance sampling.
\newblock {\em Statistics and Computing}, 31(2), 2 2021.

\bibitem{Pages}
Gilles Pag{\`e}s.
\newblock {Introduction to vector quantization and its applications for
  numerics}.
\newblock {\em {ESAIM: Proceedings and Surveys}}, 48:29--79, January 2015.
\newblock 54 pages.

\bibitem{Perrin}
T.V.E. Perrin, O.~Roustant, J.~Rohmer, Olivier Alata, J.P. Naulin, D.~Idier,
  R.~Pedreros, D.~Moncoulon, and P.~Tinard.
\newblock Functional principal component analysis for global sensitivity
  analysis of model with spatial output.
\newblock {\em {Reliability Engineering and System Safety}}, 211:107522, 07
  2021.

\bibitem{sire}
Charlie Sire, Rodolphe Le~Riche, Didier Rullière, Jérémy Rohmer, Lucie
  Pheulpin, and Yann Richet.
\newblock Quantizing rare random maps: application to flooding visualization.
\newblock {\em Journal of Computational and Graphical Statistics}, 0:1--31,
  2023.

\end{thebibliography}

\end{document}